\documentclass[preprint,12pt]{aastex}
\usepackage{graphicx}
\usepackage{setspace}
\usepackage{floatflt,multicol}
\def\1624{4U~1624$-$490}
\def\spose#1{\hbox to 0pt{#1\hss}}
\def\approxlt{\mathrel{\spose{\lower 3pt\hbox{$\sim$}}
        \raise 2.0pt\hbox{$<$}}}
\def\approxgt{\mathrel{\spose{\lower 3pt\hbox{$\sim$}}
        \raise 2.0pt\hbox{$>$}}}

\begin{document}

\title{Solid State Astrophysics \\ 
Probing Interstellar Dust and Gas Properties with X-rays \\ 
\protect{\vskip 0.3in}
{\normalsize  A White Paper Submitted to the Astro2010 Decadal Survey \\
for Astronomy and Astrophysics} }

\author{ 
	Julia C. Lee$^1$ and  
	Randall K. Smith$^2$ \\ 
	{\it with} \\ 
	C. R. Canizares$^3$,
        E. Costantini$^4$,
        C. de Vries,$^4$,
	J. Drake$^2$,
	E. Dwek$^5$,
        R. Edgar$^2$,
	A. M. Juett$^5$,
	A. Li$^6$,
	C. Lisse$^7$,
	F. Paerels$^8$,
        D. Patnaude$^2$,
	B. Ravel$^9$,
        N. S. Schulz$^3$,
	T. P. Snow$^{10}$,
        L. A. Valencic$^{11}$, 
        J. Wilms$^{12}$, 
        J. Xiang$^1$}
\affil{$^1$Harvard University   \\
	$^2$Smithsonian Astrophysical Observatory  \\
	$^3$Massachusetts Institute of Technology  \\
	$^4$SRON-NL \\
	$^5$NASA/Goddard Space Flight Center  \\ 
	$^6$University of Missouri \\
	$^7$Johns Hopkins University Applied Physics Laboratory \\
  	$^8$Columbia University \\
  	$^9$NIST \\
	$^{10}$University of Colorado\\
	$^{11}$Johns Hopkins University \\
  	$^{12} $Universit\"at Erlangen-N\"urnberg-Germany
	}

\maketitle
\pagebreak
\setcounter{page}{1}  
\vspace{-0.15in}
\begin{center}
{\bf Abstract}
\end{center}
\vspace{-0.25in}
\begin{center}
\begin{minipage}[h]{5.7in}
{\small
\begin{spacing}{0.9}
The abundances of gas and dust (solids and complex molecules) in the
interstellar medium (ISM) as well as their composition and structures
impact practically all of astrophysics.  Fundamental processes from
star formation to stellar winds to galaxy formation all scale with the
number of metals.  However, significant uncertainties remain in both
absolute and relative abundances, as well as how these vary with
environment, {\it e.g.}\ stellar photospheres versus the interstellar
medium (ISM).  While UV, optical, IR, and radio studies have
considerably advanced our understanding of ISM gas and dust, they
cannot provide uniform results over the entire range of column
densities needed.  In contrast, X-rays will penetrate gas and dust in
the cold (3~K) to hot ($10^8$~K) Universe over a wide range of column
densities ($N_{\rm H}\sim 10^{20-24} \rm cm^{-2}$), imprinting 
spectral signatures that reflect the individual
atoms which make up the gas, molecule or solid.  {\bf X-rays therefore
  are a powerful and viable resource for delving into a relatively
  unexplored regime for determining gas abundances and dust properties
  such as composition, charge state, structure, and quantity via
  absorption studies, and distribution via scattering halos. }

\end{spacing}
}
\end{minipage}
\end{center}
\vspace{-0.25in}
\begin{center}
{\bf Introduction}
\end{center}
\vspace{-0.15in}

\noindent Understanding astrophysical processes, from galaxy evolution
to star formation to stellar or AGN outflows, requires an
understanding of the 
metal abundances in the surrounding medium, in gas {\it and} dust phases.
As the primary repository of metals in the interstellar medium (ISM),
dust plays a major role in the chemical evolution of stars, planets,
and life.  Although it makes up only $\sim$1\% of the baryonic mass of
our Galaxy, it accounts for virtually all of the UV/optical
extinction, in scattering and absorption.  This extinction is so
efficient that only 1 in 10$^{12}$\ of the optical photons created in
the Galactic Center reaches us.
\begin{floatingfigure}[r]{2.8in}
\hspace{-0.3in}\includegraphics[totalheight=1.9in]{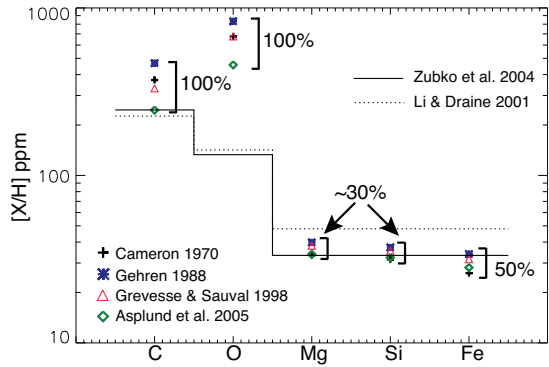}
\caption{\small \it Measured solar abundances of important metals have varied
  between 30-100\% over the last 40 years, which has had significant effects
  on the range of allowable dust models.\label{fig:solar}}
\end{floatingfigure}
In addition, ISM dust and molecules store the heavy elements needed
for making Earth-like planets, while their physics and chemistry play
an important role in giant molecular clouds, where dust acts as a
catalyst in the formation of the organic compounds vital to the
development of life \citep{Whittet03}. {\bf Accurate
measurements of the distribution of gas and dust phase abundances
in galactic environments are therefore
crucial for improving our understanding of a wide range
of astrophysical processes from nucleosynthesis to planet formation.}

Despite extensive multi-wavelength studies (primarily in radio to IR
bands), we have a limited understanding of dust grain size
distributions, their elemental and crystalline composition, and their
distribution in astrophysical environments ranging from the cold ISM
to the hot disks and envelopes around young stars and compact sources.
One immediate difficulty arises from our incomplete knowledge of the
overall material available in the gas and dust phases
(Fig.~\ref{fig:solar}). Dust abundances are traditionally inferred
from the observed depletion of the gas phase relative to the assumed
standard solar abundances (or some linear fraction of them), rather
than directly measured.  The best direct measurements of gas
abundances thus far have come from optical \& UV absorption-line
studies of diffuse interstellar clouds \citep[e.g.][]{ss96}.  However,
these results are limited to low opacity sightlines, and cannot
penetrate dense molecular clouds.  IR and radio spectroscopy can see
into dense clouds to measure many ISM molecules and compounds (mostly
polycyclic aromatic hydrocarbon or PAHs, graphites, certain silicates,
and ice mantle bands).  Yet, since these spectral features come from
probing complex molecules as a whole through rotational or vibrational
modes originating from e.g.  the excitation of phonons (rather than
electrons), they require sophisticated modelling efforts or matching
to laboratory spectra.  Reproducing the exact IR spectral features
observed has proven difficult in many cases, making robust abundance
and structure determinations for these molecules dependent upon
assumptions about their physical properties \citep{Draine03b}.

\vspace{-0.3in}
\subsection*{The power of X-ray ISM studies of dust and gas}
\begin{floatingfigure}[r]{4.2in}
\hspace{-0.2in}\includegraphics[totalheight=4.0in,angle=270]{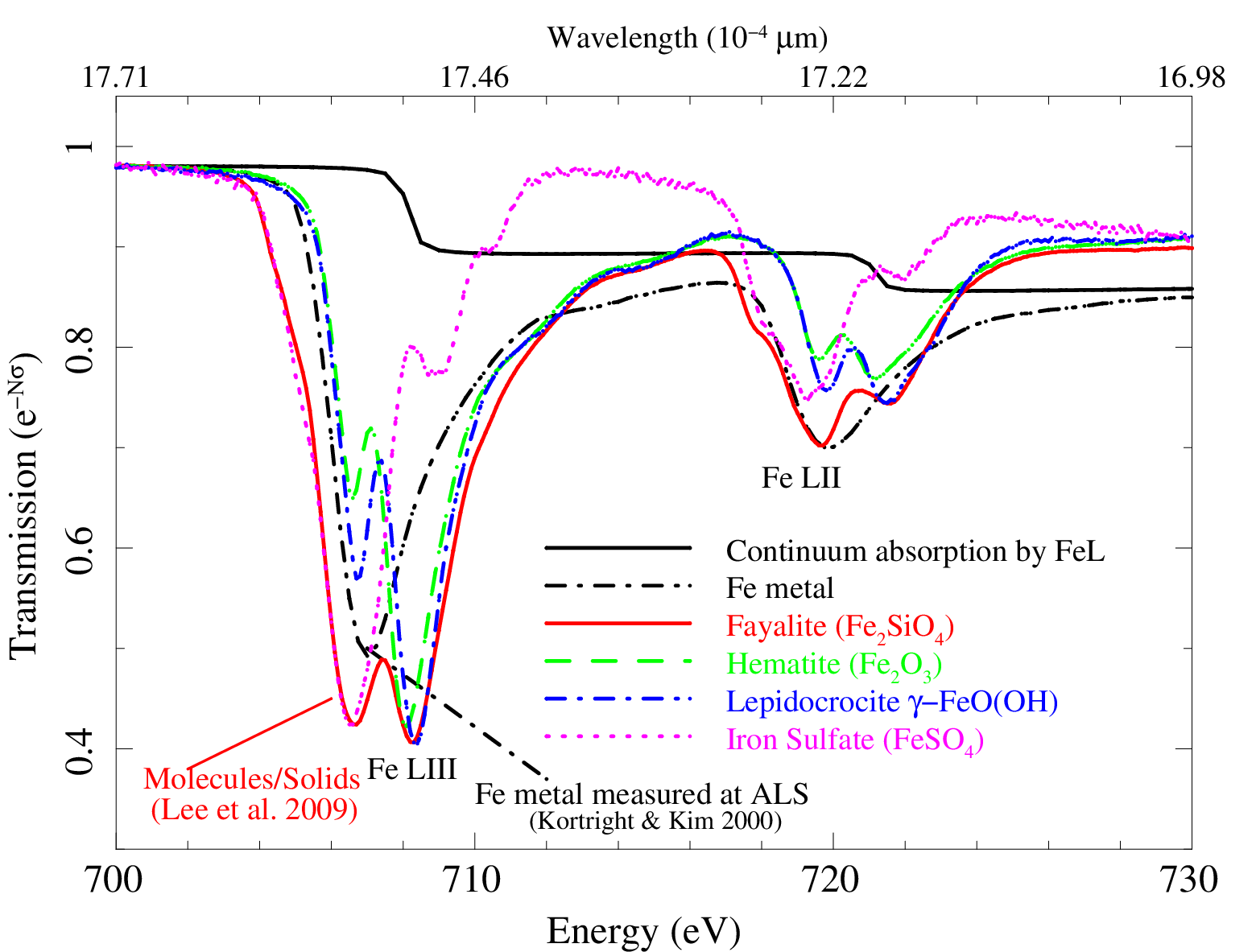}
\caption{\small \it Absorption calculated from laboratory cross
  section ($\sigma$) measurements for 0.7~keV FeL
  \citep{lee09:softxls}.  At this high spectral resolution (R=3000,
  the baseline IXO value), different molecules (color), metals
  (dashed-dot black) and continuum absorption (solid black), are
  easily distinguished.
  \label{fig:fel}}
\end{floatingfigure}
\vspace{-0.1in} X-rays probe {\it all}\, ions from neutral atoms to
hydrogenic ions, including H-like Fe~{\sc xxvi} (Fe$^{+25}$), thereby
providing a unique window into the cold (3~K) to hot ($10^8$~K)
Universe, over a wide range of column densities ($N_{\rm H}\sim
10^{20-24} \rm cm^{-2}$), not possible in any other waveband (see
\citealt{pk03} and references therein).  Detailed X-ray studies of the
ISM began with \citet{mls_crc_ism86}, who first detected photoelectric
edges along with a tentative detection of an atomic O{\sc i} line.
Fifteen years later, the Chandra and XMM gratings have facilitated
stringent ISM abundance measurements of {\it individual} ionic
species, from e.g. O{\sc i-vii}
(e.g. \citealt{xrb_paerels:01,juett_ism:04,juett_ism:06}).  To draw
similar parallels, the improved spectral resolution and throughput
available to future studies will allow us to routinely use X-ray
spectra to determine the {\it quantity and composition} of dust, where
such studies are now pushing the limits of extant satellite
capabilities (see \citealt{lee09:softxls}).  These important studies
can be facilitated as additional bonus science spawned from targeted
studies of X-ray bright objects whose lines-of-sight also intersect
with inter- and intra-stellar gas and dust with enough opacity to
imprint their unique spectral signatures.  For the remainder of this
white paper, we focus on the unique means by which X-ray studies can
be used to combine condensed matter/solid state theory and atomic
physics techniques to the study of astrophysical gas (in photoionized
and collisional environments), and dust (composition, quantity,
structure, and distribution), that is highly complementary to other
wavebands and fields (e.g. planetary science, chemistry, geology, and
experimental physics, as relevant to laboratory astrophysics).  As
such, we can expect such studies to have relevant applications which
span the Galactic to Cosmological.

\subsection*{An X-ray absorption method for determining gas-to-dust
ratios\label{sec:xafs}}
\begin{floatingfigure}[r]{3.1in}
\includegraphics[width=0.35\textwidth, angle=-90]{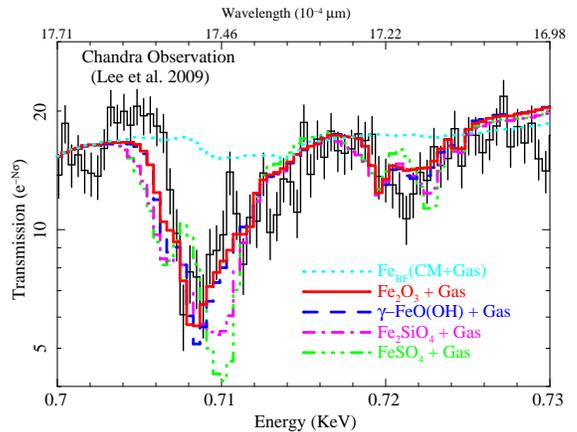}
\caption{\small \it A 15ks Chandra HETGS observation of Cygnus X-1
  (black) zoomed in on the $\sim 700$eV ($\sim 17.7$\AA) FeL spectral
  region shows that at the current highest available spectral
  resolution, we can discern at high confidence between $Fe_2O_3$
  (red) and $FeSO_4$ (green), but not between e.g. $Fe_2O_3$
  and FeO(OH) (dark blue). In light blue is continuum absorption
  by Fe~L from both gas and solids (CM).  In addition,  even for a source as bright as
  Cyg~X-1,  a 15~ks integration is require to achieve a S/N $\sim 5$\,per bin.
  Future missions with higher throughput will allow us to engage in
  such studies with more sources, and accompanying higher spectral
  resolution as e.g. that shown in Fig.~\ref{fig:fel} will enable more
  involved studies of dust structure, not currently possible.
\label{fig:cygx1}}
\end{floatingfigure}

\noindent {\it Both} gas and dust (when $\approxlt 10\mu$m ) are
semi-transparent to X-rays.  Therefore, these high energy photons can
be used to make a {\it direct measurement of condensed phase chemistry
  via a study of element-specific atomic processes}, whereby the
excitation of an electron to a higher-lying, unoccupied electron state
(i.e.  band/molecular resonance) will imprint signatures in X-ray
spectra that reflect the individual atoms which make up the molecule
or solid.  In much the same way we can identify ions via their
absorption or emission lines, the observed spectral modulations near
photoelectric edges, known as X-ray absorption fine structure (XAFS)
provide unique signatures of the condensed matter that imprinted that
signature.  Fig.~\ref{fig:fel} shows how XAFS signatures of different
molecules and solids can be identified by their unique structure and
wavelength.  With high X-ray spectral resolution, and throughput, one
can easily assess these details to determine astrophysical dust and
molecular properties.  At the current highest spectral resolution
(R=1000) of the Chandra HETGS, one can discern at high confidence
between e.g. $\rm Fe_2O_3$ and $\rm FeSO_4$ (Fig.~\ref{fig:cygx1}),
but not between $\rm Fe_2O_3$ and FeO(OH)\footnote[3]{We note that,
  while XAFS theory is quite mature, the study of XAFS is still
  largely empirical.  Therefore, the success of such a study will
  require that space-based measurements be compared with empirical
  XAFS data taken at synchrotron beamlines to determine the exact
  chemical state of the astrophysical dust.}, as would be possible at
the $R=3000$ spectral resolution of Fig.~\ref{fig:fel}.  Furthermore,
higher throughput and spectral resolution than currently available
will provide additional important knowledge about grain structure
(i.e. bond lengths and charge states).  Therefore, X-rays should be
considered a powerful and viable resource for delving into a
relatively unexplored regime for determining dust properties:
composition, charge state, structure, quantity (via absorption
studies; see \citealt{lee09:softxls} for discussion on how to
determine quantity {\it and} composition), and distribution (via
scattering halo studies; e.g. \citealt{xiang07:halo} and references
therein; details to follow).

Chandra and XMM have begun this work but have been limited by the
available effective area at high spectral resolution
(e.g. \citealt{jclmcg6wa1:01,jcl_grs1915:02,ism_takei:02,ism_ueda:04, 
scox1xafs}).  As such, {\it condensed matter astrophysics}
(i.e. the merging of condensed matter and astrophysics
techniques for X-ray studies of dust) is now possible, albeit difficult
due primarily to the need for more S/N.   As stated, for more involved
studies of grain structure, higher spectral resolution than currently
available is also needed.

To illustrate the complementarity of X-ray studies with those in other
wavebands, consider \citet{lee09:cygx1xafs}'s Spitzer IR and X-ray
study of the line-of-sight toward the BH binary Cygnus~X-1 which shows
different grain populations:~IR detections of silicates versus X-ray
studies ruling out silicates. This brings to bear intriguing questions
relating to grain sizes and origin: e.g. does the IR probe a
population of {\it large} silicate grains that are opaque to the
X-rays, or are we probing truly different origins, or is it something
more complex?  Another example arises from Spitzer~IRS studies of
the star HD~113766, which shows tantalizing evidence for dust
associated with terrestrial planet formation \citep{Lisse08}.  
Analysis of the IR spectra provided knowledge of the mineralogy, but
could not determine overall abundance ratios, e.g. the amount of Mg vs
Fe depleted into grains, to better than a factor of 2.  This
information can be extracted from X-ray studies of XAFS spectra with sufficient 
S/N and spectral resolution.  
These examples highlight the
power of an orthogonal approach using X-rays to complement other
wavebands:~(1)~by probing dust properties missed by other
detection techniques, and (2)~by being sensitive to the range of
sub-micron to $\sim$10~$\mu m$ size dust in $N_{\rm H} \sim 10^{20-24}
\rm cm^{-2}$ environments.
Such multi-wavelength studies, especially if achieved with high throughput 
and spectral resolution in the X-ray band, will add significantly
to our knowledge of the mineralogy and distribution of nascent and
interstellar dust populations and therefore the environments from
which they originate, be they outflows or star forming regions.

\begin{figure}[t]
\begin{center}
\includegraphics[totalheight=1.8in]{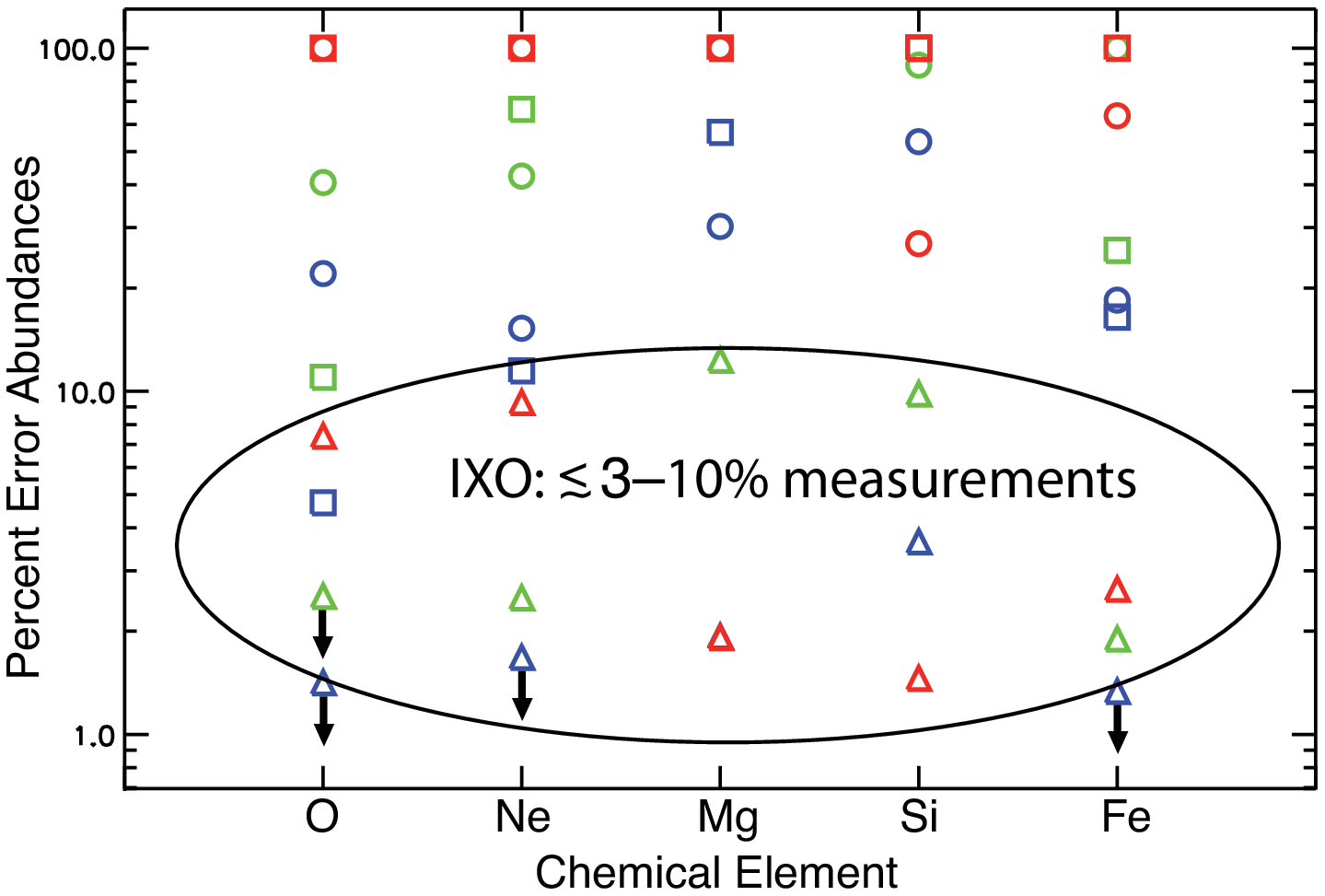}
\includegraphics[totalheight=1.7in]{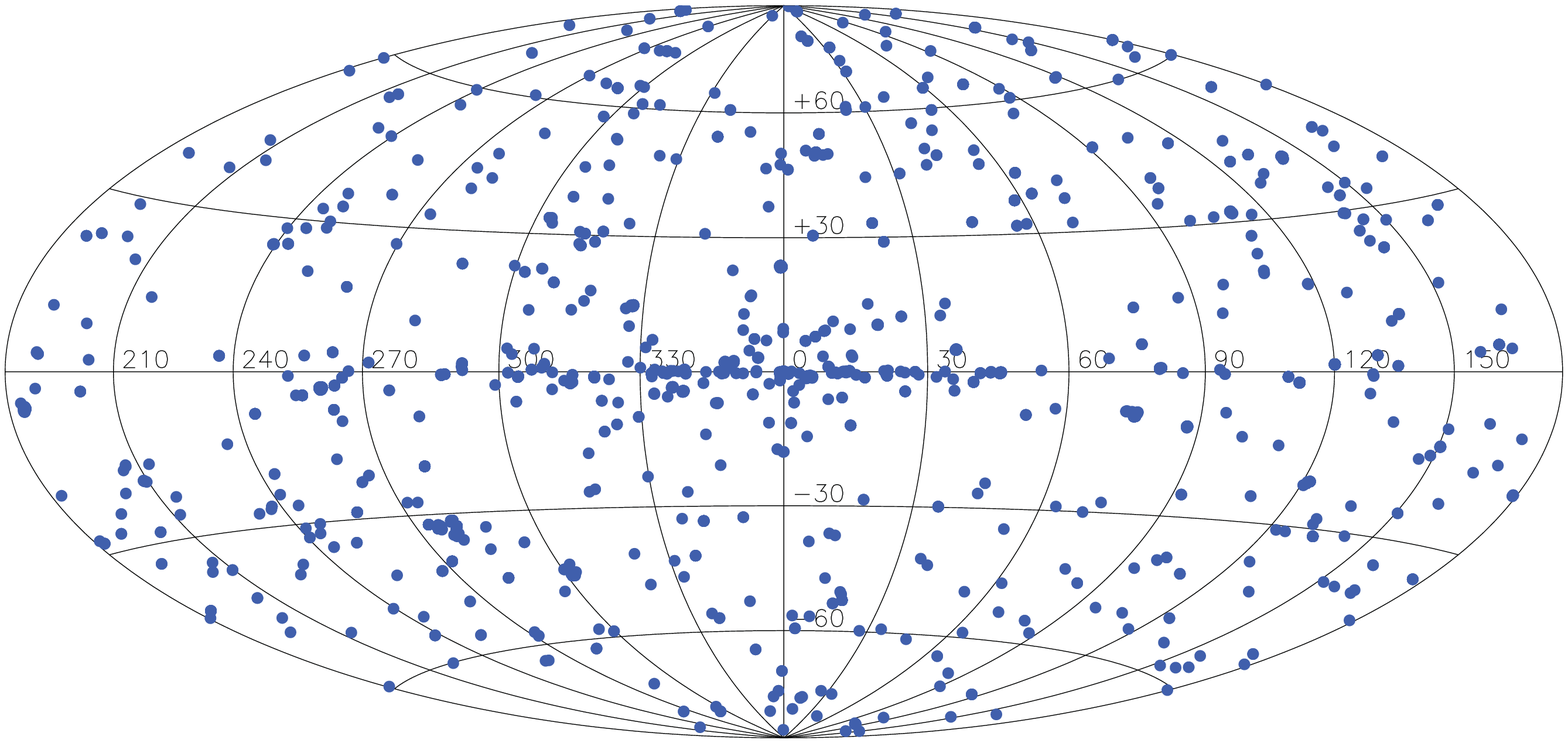}
\end{center}
\caption{\small \it [Left] Potential abundance measurement accuracy
  from spectra with 50 ksec exposures using the Chandra ACIS-S, LETG
  (circles), XMM-Newton RGS (squares) and IXO calorimeter (triangles)
  for N$_{\rm H}$\ values of $5\times10^{20}$\,cm$^{-2}$\ (green),
  $5\times10^{21}$\,cm$^{-2}$\, (blue), and
  $5\times10^{22}$\,cm$^{-2}$\,(red). [Right] Sightlines to bright AGN
  or X-ray binaries for which IXO will be able both study the source
  and determine abundances to better than 10\%. \label{fig:sightlines}}
\end{figure}

Despite the power of using X-rays for dust studies, this energy band
has not been fully exploited for spectral studies of gas, dust and
molecules, due largely to the unavailability of instruments with {\it
  both} good throughput and spectral resolution ($R \approxgt 1000$).
Missions such as IXO can push the frontiers of such studies by
enabling high-precision elemental abundance measurements of {\it gas
  and dust} towards hundreds of sightlines, both in our Galaxy and
beyond (see Figure~\ref{fig:sightlines}[Right]).  Simulations based on
IXO's target spectral resolution and area show that a S/N=10~per bin
can be achieved for over 200 sources in 30 ksec or less, while the
higher spectral resolution will allow us to realize the
previously-discussed studies of dust structure that are not currently possible.
 
IXO's energy coverage will allow us to study in detail photoelectric
edges near C~K, O~K, Fe~L, Mg~K, Si~K, Al~K, S~K, Ca~K, and Fe~K, and
therefore all gas-phase as well as molecules/grains containing these
constituents, covering all the important species with high depletion
rates onto dust.  These studies, when applied to different
astrophysical environments, Galactic or extragalactic, will enable
probes of the gamut of outstanding issues ranging from NS and BH
(stellar and supermassive) evolutionary histories, to the hot
accretion flow in dust enshrouded accretion systems (e.g. Sgr~A$^{\ast}$
and similar AGN and star formation systems; see
\citealt{lee09:softxls} for discussion), to cosmological implications
(e.g. Type Ia SNe light curves are affected by line-of-sight dust),
all using X-rays.

X-ray spectra can also determine abundances in different ISM
environments \citep{yw08ism}, thereby opening a window on the study of
grain evolution and cycling between diffuse and dense or dark clouds.
The current uncertainties for Chandra and XMM abundance measurements
usually exceed 20\%, due primarily to the low effective area of the
spectrometers.  As a result, these results do not constrain ISM
abundances with more accuracy than using stellar photospheres or
UV/optical data.
Over 1800 sources (Fig.~\ref{fig:sightlines}[Right]) have high enough
X-ray fluxes ($\approxgt 5\times10^{-12}$\,ergs cm$^{-2}$s$^{-1}$) for IXO to
make high precision (3-10\%) O, Mg, Si, and Fe abundance
determinations for both diffuse and dense sightlines, including
molecular regions where N$_{\rm H} > 5\times 10^{22}$\,cm$^{-2}$\ (see
Figure~\ref{fig:sightlines}[Left]). Along the less dense sightlines
with X-ray and UV-bright sources, UV/optical gas-phase
abundance measurements can be combined with X-ray determinations of the total
abundances to extract the ratio of gas-to-dust directly
\citep[e.g.][]{Cunningham04}.  X-ray observations will also probe
abundances beyond the Milky Way: between 10-100 ultra-luminous X-ray
sources (ULXs) have X-ray fluxes that will allow robust abundance
measurements beyond our Galaxy.  With spectral resolution $R>1000$,
the Galactic and distant contributions can be separated simply using
known velocity separations.

With the additional XAFS studies of dust composition possible for over
200 sources along different sightlines, we can also discriminate
between different grain models.  With these many sight-lines, we will
obtain valuable information on the chemical uniformity of the
ISM, including mixing and enrichment.

\subsection*{Dust-induced X-ray Scattering\label{sec:halo}}
\begin{floatingfigure}[r]{3.2in}
\includegraphics[totalheight=2.0in]{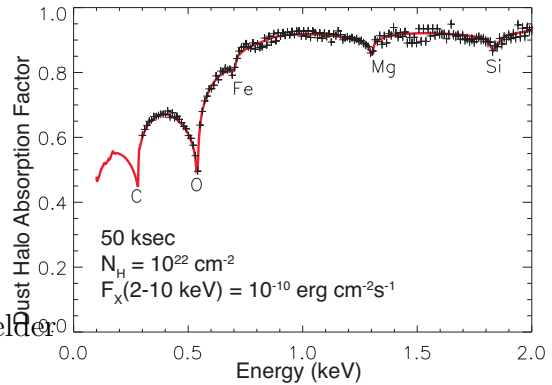}
\caption{\small \it A  50 ksec IXO simulations showing the effects of
  different elemental composition on scattering halo properties as a
  function of energy.  This will allow the abundances in dust grains
  as a function of grain size to be measured directly from the scattered halo.
 \label{fig:shalo}}
\end{floatingfigure}

Scattered X-rays also provide information on dust
grain sizes, positions, and {\it composition}, via {\it imaging}
studies of the arcminute-scale halos around X-ray bright sources
created by small-angle scattering in interstellar dust grains \citep{Draine03}.
In some cases, we can even use such studies to more accurately
determine distances to compact objects or even other galaxies (see
\citealt{trumper73} for a description of the method;
\citealt{predehl00, db04, xiang07:halo} for applications).  Measuring
the location and size of grains -- seen most dramatically in the
dust-scattered rings around X-ray bursts \citep{Vaughan04} -- provides
information about the grain environment and their formation processes
when combined with composition and abundance studies discussed
previously.

The intensity of the halo created by small-angle scattering of X-rays
by dust is especially sensitive to the large end of the grain size
distribution, as these grains dominate the scattering cross section
($\sigma \propto \rho^{2} a^{4}$).  In addition, every IXO
observation of 20 ksec or longer of a moderately bright source
($>$10$^{-11}$ erg s$^{-1}$ cm$^{2}$) with N$_{\rm H} \approxgt
10^{21}$\,cm$^{-2}$\ will contain a detectable X-ray dust halo.  By
measuring the total {\it scattered} halo intensity {\it as a function
  of energy} in bright sources -- {\it absorption}\ features from the
elemental grain composition can be measured \citep{Costantini05}, as
shown in Figure~\ref{fig:shalo} for an IXO simulation.  {\bf These
  X-ray halos will directly measure the composition of the large
  grains.}

Existing interstellar grain size distributions have been developed
through information gathered from UV/optical/NIR extinction,
polarization, IR emission, and depletion.  By combining measurements
towards hundreds of sight-lines, modelers have constrained the
composition, shape, and grain size distributions from tens of \AA~up
to $\sim$0.5~$\mu$m; above this size, grains act as size-independent
``gray'' particles in these wavebands.  Despite the lack of detailed
information, the models agree that even in the diffuse ISM, grains
larger than 0.1$\mu$m contain most of the dust mass in the ISM.
Moreover, most interstellar dust resides in dense molecular clouds,
where UV/optical observations are difficult to impossible.

Early interstellar grain models used an approach consisting of a
population of bare silicate and graphite grains in a simple power law
size distribution, designed to reproduce the ``average'' Galactic UV
extinction curve \citep{mrn77}.  Since that time, many workers have
developed models that were constrained by data from different
wavelength regimes, as more information became available
\citep[e.g.][]{Dwek97, LG97, LD01, wd01, ZDA04}.  However, a mismatch
of up to $\sim30\%$\ between the total grain mass and the available
metals remains in many models (see Figure~\ref{fig:solar};
\citealt{Draine03b, ZDA04}).  It is not yet clear if the solution will
involve changes to the ISM abundances or to the grain models.  Halos
seen with Chandra and XMM have already shown that adjusting the grain
porosity does not solve the problem \citep{Smith02, Smith08, VS08}; more
accurate abundance data are sorely needed.

Amongst other advantages, X-ray studies will afford us a window into
giant molecular clouds, which are opaque to UV and optical light.
While grains may coagulate there and/or grow envelopes of ices
and organic material \citep[e.g.][]{CM88,Vrba93, Whittet01},
what little we know about these chemical warehouses come from observed
molecular transitions in the IR and microwave, which are primarily
sensitive to ice and organic compounds.  Theoretical work by
\citet{CL05} offer that far-IR and sub-millimeter polarization studies
may shed light on grain sizes in dark clouds (A$_{\rm V} > 10$),
although this remains to be demonstrated.  By capitalizing on the
well-known sensitivity of X-ray scattering halos on grain sizes,
X-rays can be used to provide, again, an orthogonal means by which we can
assess dark cloud cores to compare with studies in the IR.  The halos
of bright objects behind these clouds will let us examine the grain
growth processes which heretofore have been shrouded in mystery, while
the observed fine structure near photoelectric absorption edges
discussed above will provide valuable information on the
composition and growth of organic mantles.

\begin{multicols}{2}

\end{multicols}
\newpage
\end{document}